# LATTICE BOLTZMANN SIMULATIONS OF SELF-PROPELLING CHIRAL ACTIVE DROPLETS

**Livio Nicola Carenza[1]\*, Giuseppe Gonnella[1] and Giuseppe Negro[1\*]**

1: Dipartimento di Fisica, Università degli studi di Bari and INFN, via Amendola 173, Bari, I-70126, Italy

**Abstract** *Active matter describes materials whose constituents are driven out of equilibrium by continuous energy consumption, for instance from ATP. Due to the orientable character of the constituents, active suspensions can attain liquid crystalline order and can be theoretically described as active liquid crystals. Their inherently nonequilibrium dynamics causes a range of new striking effects, that in most cases have been characterized with numerical simulations, using lattice Boltzmann models (LB).*

*In many active biological systems chirality plays an important role. Biomolecules such as DNA, actin or microtubules form helical structures which, at sufficiently high density and in the absence of active forces, tend to self-assemble into twisted cholesteric phases. Understanding the outcome of the interplay between chirality and activity is therefore an important and timely question. Studying a droplet of chiral matter in 3D, we have found evidence of a new motility mode, where the rotational motion of surface topological defects, that arrange in a fan-like pattern. The resulting regular propulsive motion due to the underlying chirality is a striking phenomenon that can be also used in practical applications. The use of a parallel (MPI) implementation of lattice Boltzmann models, and available HPC resources, have been of fundamental importance in conducting the study. We have used different HPC clusters, and among these RECAS. This allowed us to conduct a scaling test performed on the different computational infrastructures.*

# 1. INTRODUCTION

Active matter comprises condensed matter systems where the constituent units consume internal energy depots to move or to do some work on their environment [1]. Therefore, active systems are inherently out of thermodynamic equilibrium. Living systems, groups of animals, flock of birds, schools of fish, swimming bacteria, can be classified as active matter systems [2-4]. An example at microscopic scales can help to better understand the difference with respect to usual equilibrium systems. Consider Brownian colloidal particles (pollen or mineral grains in the original experiments by R. Brown [5]) suspended in a fluid. These particles, showing a random diffusive motion, share with the surrounding fluid the important property of having the same temperature, which in the case of Brownian particles is represented by the mean kinetic energy. The equipartition property of energy applies in this case. Think now about living units (bacteria, algae, etc.) in a solution. One would observe a behavior given by a combination of diffusion with ballistic motion [1]. The last is due to intrinsic self-propulsion forces acting in some predefined directions, for example the main geometrical axis in the case of not symmetric swimmers. These systems will operate at a temperature, a quantity to be also properly defined, different from that of the fluid bath, and are characterized by a continuous production of entropy [6].

Beyond the interest in extending traditional thermodynamics to the description of these systems, active matter is also quite relevant for applications. Inspired by real biology, artificial systems have been created with similar motility properties as those of living swimmers [7]. Artificial active colloids show coherent unidirectional motion in small channels, form small aggregates also in absence of attractive interactions, and can be used for drug delivery or micromotors [8]. Other classes of active fluids also show interesting properties. In this article we will be mainly concerned with a class of systems, sometimes called active liquid crystals [9], where the active units are organic filaments, obtainable by cellular extracts, coupled to molecular motors [10]. At single or few units level, these molecules exert a stress field corresponding to extensile (for microtubule filaments) or contractile (for acto-myosin) local flows [11]. An important question is if the local flow can be converted in macroscopic flow, with some applicative relevance. Previous studies of active liquid crystals have shown the occurrence of peculiar rheological properties [12,13]. For example these systems can flow with an effective zero viscosity in a regime that resembles superfluid behavior [14]. Experimental realizations of active liquid droplets have been done in the laboratories of Sagués [15] and Dogic [10].

In this article, after briefly reviewing fundamental spontaneous flow instability properties of active nematic liquid crystals, we will show that the chiral character of active filaments confined in droplets, is essential for their linear propulsion [16]. Chirality is a generic feature of most biological matter [17,18]. A microtubule–motor mixture breaks chiral symmetry in two ways. First, microtubules are intrinsically helical [19]. Second, kinesin or dynein motors exploit adenosine triphosphate hydrolysis to twist their long chains and apply a nonequilibrium active torque on the fibers they walk along [20]. We will show that the dynamics of topological defects is crucial for determining the self-propelling behavior of the droplets.
Our results are based on a generalization of continuum description of dynamics of liquid crystals simulated by Lattice Boltzmann (LB) Methods [16]. The theoretical model and the numerical methods will be respectively described in Section 2 and 3. The possibility of performing 3D simulations largely uses the efficient parallel portability of Lattice Boltzmann codes. Details will be given in Section 3 with some results for the scalability properties of the codes we used in our simulations, resulting from tests performed on different computing infrastructures. Section 4 will contain some of our results regarding the behavior of self-propelling chiral droplets.

## 2. MODEL

In this Section we will introduce the physical quantities crucial for a correct description of a droplet of Liquid Crystal (LC) embedded in an isotropic background.

We make use of a scalar concentration field $\varphi$ to distinguish between the interior of the droplet, where the LC is confined ($\varphi > \varphi_{cr} = 1$) and an isotropic background ($\varphi = 0$). The liquid crystal order is described by the tensor field $Q_{\alpha\beta}$. This can be written as the traceless tensor product of a unit vector $\boldsymbol{n}$, accounting for the local orientation of the constituents, multiplied by a scalar parameter $S$, ranging from 0 in the isotropic phase to 1 in the regions where the order is maximum and the constituents are perfectly aligned to each other [21]. $\boldsymbol{v}$ is the velocity of the fluid, flowing under incompressible conditions, so that the density $\rho$ is constant and is not a dynamical observable [22].

The concentration field is a conserved quantity and its evolution is ruled by a convection-diffusion equation:

$$d_t \varphi = M \nabla^2 \mu, \qquad (1)$$

where $d_t$ denotes the material derivative, $M$ is the mobility and $\mu = \delta F/\delta \varphi$ is the chemical potential with $F$ a suitable free-energy (see later). The right-hand term acts as a thermodynamic force that drives the system towards equilibrium. The liquid crystal follows a relaxing evolution governed by the Beris-Edwards equation, given by

$$\tilde{d}_t Q_{\alpha\beta} = -\frac{1}{\Gamma} H_{\alpha\beta}. \qquad (2)$$

Here $\tilde{d}_t$ is the material derivative for a tensor field, including both an advective and a strain-rotational term –that have respectively the effect of dragging the LC and rotating it according to the flow direction. On the right-hand side of Eq. (2) $\Gamma$ is the rotational viscosity and $H_{\alpha\beta} = \delta F/\delta Q_{\alpha\beta}$ is the molecular field driving the system towards equilibrium. The fluid momentum follows the Navier-Stokes equations:

$$\rho d_t v_\alpha = \partial_\beta \left( \sigma_{\alpha\beta}^{pass} + \sigma_{\alpha\beta}^{act} \right). \qquad (3)$$

Here, the stress tensor is divided in a passive contribution $\left(\sigma_{\alpha\beta}^{pass}\right)$ and an active one $\left(\sigma_{\alpha\beta}^{act}\right)$. In particular, the passive contributions are due to hydrodynamic effects (isotropic pressure and viscous interactions) and to reactive phenomena arising from the mutual interaction between the fluid and the order parameters $\varphi$ and $Q_{\alpha\beta}$.

The active stress tensor has instead phenomenological origin and it aims at describing the effect of the swimming of the active particles on the surrounding fluid. A minimal (and successful) approach is to model the active agents as force dipoles [11]: by summing the contributions from each force dipole and coarse-graining, it is possible to show that the stress exerted by the active particles has

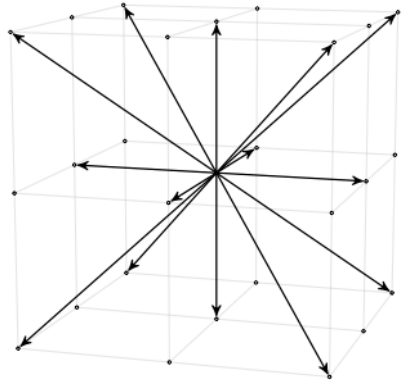

Figure 1. Representation of D3Q15 lattice. Here the black dots represent the lattice sites, while the black arrows represent the lattice velocities.

the form

$$\sigma_{\alpha\beta}^{\text{act}} = -\zeta S\left(n_\alpha n_\beta - \frac{1}{3}\delta_{\alpha\beta}\right) = -\zeta Q_{\alpha\beta}, \tag{4}$$

where $\zeta$ is a phenomenological parameter that measures the activity strength. $\zeta$ is positive for outwarding force dipoles (namely extensile swimmers or pushers) and negative for inwarding force dipoles (contractile swimmers or pullers). In this paper we will only consider the case of extensile systems.

In order to define the equilibrium properties of the system we define the free energy functional $F[\varphi, Q_{\alpha\beta}]$ which depends on the configuration of the two order parameters $\varphi$ and $Q_{\alpha\beta}$ and is given by

$$\begin{aligned}F[\varphi, Q_{\alpha\beta}] = \int dV \Bigg[&\frac{a}{2}\varphi^2\left(\frac{\varphi-\varphi_0}{\varphi_{cr}}\right)^2 + \frac{k_\varphi}{2}(\nabla\varphi)^2 \\ &+ A_0\left[\frac{1}{2}\left(1-\frac{\chi(\varphi)}{3}\right)\boldsymbol{Q}^2 - \frac{\chi(\varphi)}{3}\boldsymbol{Q}^3 + \frac{\chi(\varphi)}{4}\boldsymbol{Q}^4\right] \\ &+ \frac{K_Q}{2}[(\nabla\cdot\boldsymbol{Q})^2 + (\nabla\times\boldsymbol{Q}+2q_0\boldsymbol{Q})^2] + W\nabla\varphi\cdot\boldsymbol{Q}\cdot\nabla\varphi\Bigg].\end{aligned} \tag{5}$$

The first term, multiplied by the phenomenological constant $a > 0$, describes the bulk properties of the fluid; it is chosen in order to create two free-energy minima, one ($\varphi = 0$) corresponding to the passive material and the other one ($\varphi = \varphi_0$) corresponding to the active phase. The interfacial tension between the passive and active phase is controlled by the positive constant $k_\varphi$. The bulk properties of the LC are determined by the term proportional to $A_0$. In particular the LC phase is confined in those regions where $\chi(\varphi) = \chi_0 + \chi_s\varphi > 2.7$ [20]. The gradient terms in $K_Q$ account for the energy cost of elastic deformations, while the term proportional to $q_0$ favours twisting deformations of the LC thus introducing chirality at equilibrium level. The last term sets the anchoring of the LC at the droplet interface. In particular, tangential anchoring is achieved for $W > 0$.

## 3. LATTICE BOLTZMANN METHODS FOR COMPLEX AND ACTIVE FLUIDS

In this Section we will briefly present the lattice Boltzmann method [23], that is the numerical approach which we made use of to integrate the dynamical equations ruling the evolution of the system previously introduced.

The lattice Boltzmann approach is based on a phase-space discretized form of the Boltzmann equation for the distribution function $f(\boldsymbol{r}, \boldsymbol{\xi}, t)$ describing the fraction of fluid mass at position $\boldsymbol{r}$ moving with velocity $\boldsymbol{\xi}$ at time $t$. The algorithm can be thus expressed in terms of a set of discretized distribution functions $\{f_i(\boldsymbol{r}_\alpha, t)\}$, defined on each lattice site $\boldsymbol{r}_\alpha$. These are related to a discrete set of lattice speeds $\{\boldsymbol{\xi}_i\}$ labelled with an index $i$ which varies from 1 to $N$ according to the order of accuracy of the numerical scheme. In the scope of this work we made use of the D3Q15 lattice, namely a three-dimensional square lattice characterized by $N = 15$ lattice speeds (see Fig. 1).

The evolution of the distribution functions is governed by

$$f_i(\boldsymbol{r}_\alpha + \boldsymbol{\xi}_i \Delta t, t + \Delta t) = f_i(\boldsymbol{r}_\alpha, t) - C(\{f_i\}, t) \tag{6}$$

where $C(\{f_i\}, t)$ is the collisional operator that drives the system towards equilibrium. In the BGK approximation [24] this can be written as

$$C(\{f_i\}, t) = -\frac{1}{\tau}\left(f_i - f_i^{eq} + \Delta t F_i\right) \tag{7}$$

Here, $F_i$ is a forcing term, while the $f_i^{eq}$ are equilibrium distribution functions and $\tau$ is the relaxation time, which is related to the speed of sound and to the viscosity of the fluid.

The mass and the momentum density are defined as follows:

$$\begin{aligned}\rho &= \sum_i f_i \\ \rho v_\alpha &= \sum_i f_i \xi_{i\alpha}\end{aligned} \tag{8}$$

where summations are performed over all discretized directions at each lattice point. The same relations hold for the equilibrium distribution functions, to ensure mass and momentum conservation.

To recover the Navier-Stokes equation (Eq. (3)) in the continuum limit –that is when both the lattice spacing and the time step tend to 0– it is necessary to impose suitable conditions on the moments of both the equilibrium distribution functions:

$$\sum_i f_i \xi_{i\alpha} \xi_{i\beta} = -\sigma_{\alpha\beta}^{symm} + \rho v_\alpha v_\beta \tag{9}$$

and the forcing terms:

$$\begin{aligned}\sum_i F_i \xi_{i\alpha} &= 0 \\ \sum_i F_i \xi_{i\alpha} &= \partial_\beta \sigma_{\alpha\beta}^{asymm} \\ \sum_i F_i \xi_{i\alpha} \xi_{i\beta} &= 0\end{aligned} \tag{10}$$

Conditions (8) and (9) can be satisfied by expanding the equilibrium distribution function up to the second order in the fluid velocity [25]:

$$f_i^{eq} = A_i + B_i(\boldsymbol{\xi} \cdot \boldsymbol{v}) + C_i v^2 + D_i(\boldsymbol{\xi} \cdot \boldsymbol{v})^2 + \boldsymbol{G}_i : \boldsymbol{\xi} \otimes \boldsymbol{\xi}. \tag{11}$$

We observe that this expansion is valid as far as the typical velocity of the fluid is small compared with the speed of sound. The coefficients appearing in the expansion can be calculated by means

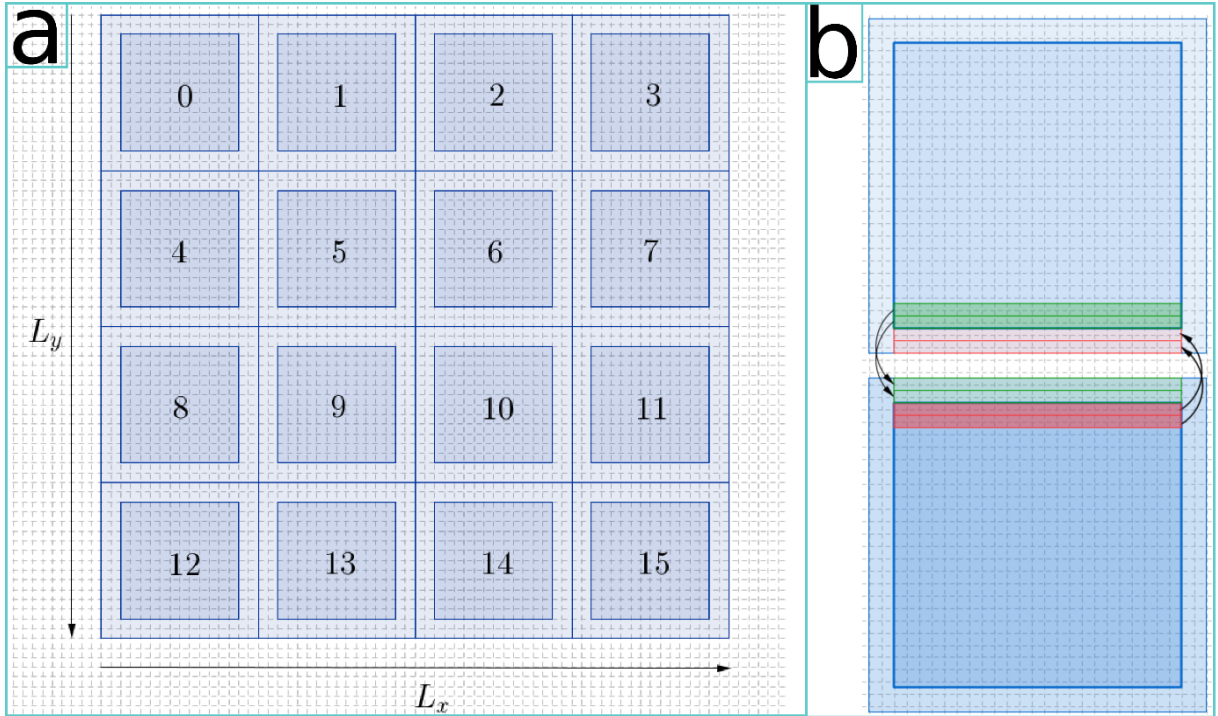

Figure 2. Panel (a) shows the division of the physical grid in subdomains, each of whom is assigned to a computational unit in the MPI communicator. Each subdomain is framed with *ghost cells* here denoted in light blue. Panel (b) shows how physical data are exchanged among two neighboring processors. Physical data stored in the dark green lattice sites by the upper processor are sent to the bottom processor and placed in the light green ghost cells. The same procedure applies with the physical data stored in the dark red lattice sites. At the end of the process each processor has a copy of the data stored in the memory position of the neighboring processor.

of Eq. (8) and (9).

The dynamics of the two order parameters $\varphi$ and $Q_{\alpha\beta}$ has been integrated by means of a predictor-corrector finite difference algorithm implementing a first-order upwind scheme and fourth order accurate stencils for space derivatives [26].

### 3.1. Computational features

In this Section we will comment on the performances of lattice Boltzmann algorithms, compared to other numerical schemes commonly implemented for solving the Navier-Stokes equations.

In particular, lattice Boltzmann methods are computationally efficient and suitable for parallel approaches. For instance, methods such as finite-difference (FD) and pseudo-spctral (PS) algorithms require high order of precision to ensure stability and to correctly compute non-linearities in the NS equation [26]. This introduces non-local operations in the computational implementation that reduce the throughput of the algorithm. On the contrary, lattice Boltzmann algorithms are intrinsically local. This is because the interaction between lattice sites is restricted to first and second neighbors defined by the lattice speeds. Indeed, integrating the hydrodynamic equations on a d-dimensional cubic grid of size L by means of a pseudo-spectral method requires a number of floating operations $\sim (\ln L) L^d$, while a LB approach only requires $\sim L^d$ [27]. Nevertheless, lattice Boltzmann methods may be extremely memory consuming, especially in three-dimensional geometries. This is because it is necessary to store the values of both the distribution functions and the velocity field for each lattice site, while in other numerical approaches storing the only velocity field is enough. To solve such issues and to speed up simulation times, one may think to make use of a parallelization approach, in which the full computational problem is split among different tasks, that can be treated independently by different processors. This approach proves to be an excellent solution for the two typical issues in computational physics: execution time and memory allocation. Indeed, having more computational units working on the same problem at the same time allows for exploiting their memory resources and for a drastic reduction of the simulation times (thus allowing for the exploration of larger systems or improving the accuracy of the simulation).

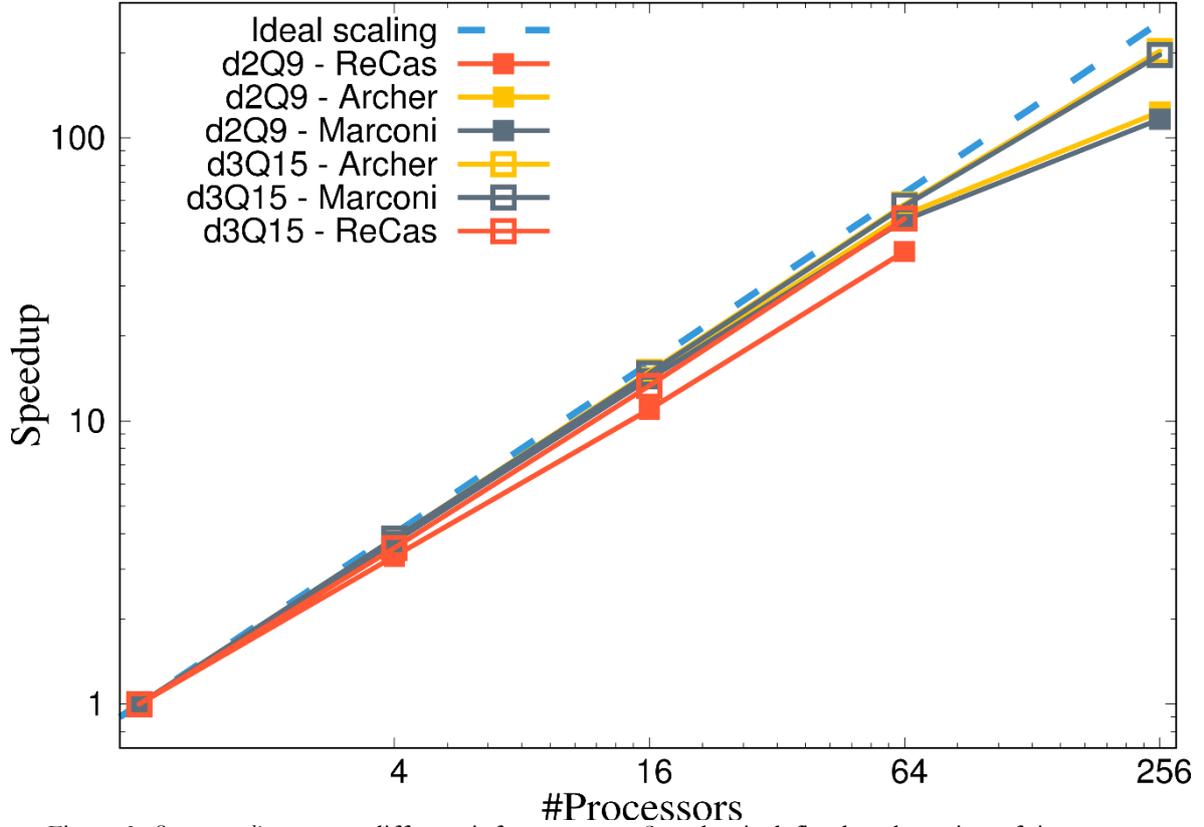

Figure 3. *Strong scaling* test on different infrastructures. Speedup is defined as the ration of time spent to perform the simulation with only one processor over the time taken when more processors are used. Simulations were performed in 2D on a square lattice of size 512x512 and in 3D on a cubic domain of size 128x128x128. We ran the test on different HPC farms: Archer UK National Supercomputing Service (Edinburgh - *www.archer.ac.uk*), CINECA Marconi – Skylake partition (Casalecchio di Reno (BO) - *www.hpc.cineca.it*) and ReCas – Bari (*www.recas-bari.it*).

Among the possible parallel approaches, we chose to parallelize the code by means of MPI –a protocol allowing different processors to communicate among themselves and share information at runtime. The idea is to split the computational grid in subdomains to be assigned to a different computational unit. In line of principle, this would allow for a speed-up equal to the number of processors taking part to the MPI communicator. Nevertheless, a number of issues may influence the scalability of the code. Indeed, even if LB algorithms are essentially local, still they need to compute derivatives and perform the streaming of the distribution functions at each iteration. This becomes an issue for the lattice sites on the edge of the subdomains, since some of the relevant variable may be stored in a memory position not accessible by the processing unit. One possible approach to overcome this issue is the *ghost-cell method* consisting in allocating extra lattice sites at the side of the physical subdomains whose purpose is to store the side configuration of the neighboring domains (see Fig. 2). Thus, processors can fill in the ghost cells by sharing information via the *send-receive* query in the MPI protocol. Processor synchronization and latency times related to the architecture of the computing environment may then slow down the execution.

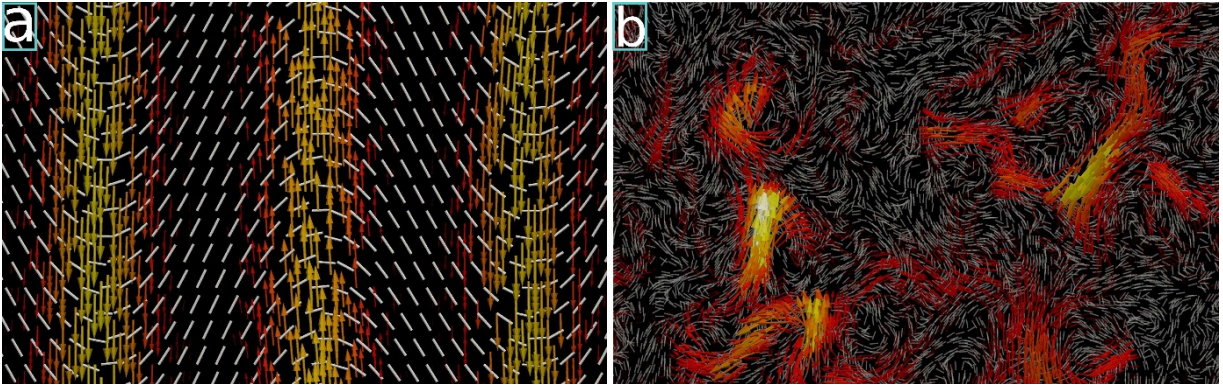

Figure 4. Different regimes of an extensile ($\zeta > 0$ in Eq. (4)) active nematic LC. (a) For low activity the active nematic LC relaxes into a *quiescent* state, where the LC is globally ordered and flows are almost suppressed. (b) If the intensity of active forcing is raised, the liquid crystal which models the active components becomes unstable and settles in a wavy pattern. Bending deformations of the nematic pattern act as a source of momentum thus generating a self-sustained and stationary flow here represented with arrows colored according to the intensity of the flow (yellow corresponds to the maximum). (c) By further increasing activity, the liquid crystal deformations strengthen, generating more and more intense flows that drive the system into a chaotic state (panel c of Fig. 4). The typical flow pattern resembles those observed in turbulent low viscosity media.

Fig. 3 shows the results of a *strong scaling* test performed on the code integrating the hydrodynamic equations (1)-(3). This test consists in changing the amount of processors used to perform a certain task, while keeping fixed the size of the computational grid and measuring the speedup, namely the ratio of time spent to perform the operation with only one processor over the time taken when more processors are used. Simulations were performed both in 2D on a D2Q9 lattice (hollow dots) and in 3D on a D3Q15 (full dots) lattice structures on different computational infrastructures (Archer (red), Marconi (blue) and ReCas (green)). While for a few number of processors the scaling is approximately linear, thus close to the ideal linear behavior (black line), as the number of processors increases, it progressively deviates from the ideal scaling law. As stated before, this is due to a number of issues that may depend both on the infrastructure characteristics (bandwidth, cache size, latency, etc.) and on the program implementation (bottlenecks, asynchrony among processor, etc.). Moreover, code scalability is found to be significantly better in three-dimensional grids than in their bidimensional version. This is because the fraction of time spent by the 3D code to perform parallel operations (sending and receiving data, reduction operations, synchronization, etc.) is consistently reduced compared with its 2D counterpart. These features are not significantly influenced by the particular infrastructure that we considered –a result that carries to the important conclusion that the scalability property of the three firms are comparable.

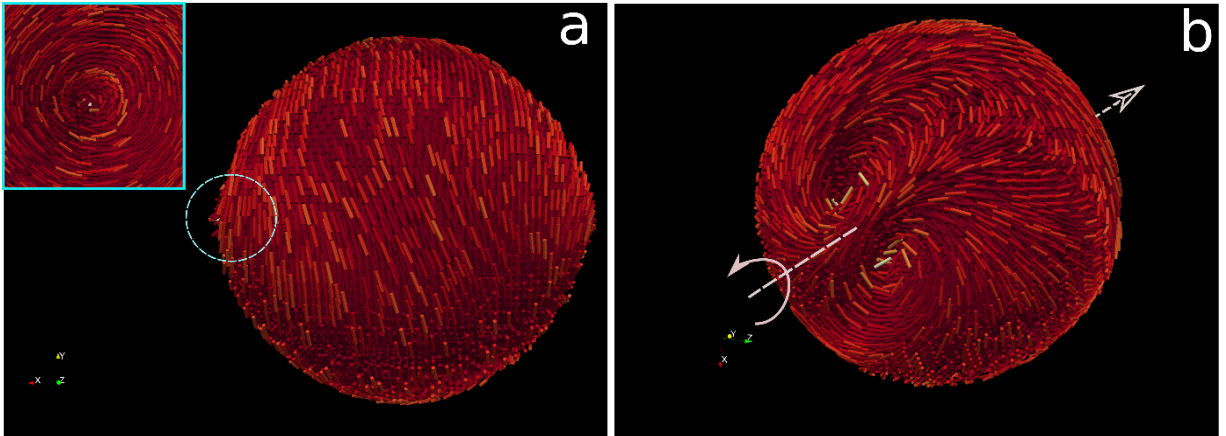

Figure 5. Different regimes of an extensile active cholesteric LC. Both the panels show the director field on the surface of the droplet, where tangential anchoring requires the formation of two vortical defects. Their position depend on both the strength of chirality (equal in both panels) and activity. In particular, panel (a) shows the case at small activity, where two antipodal aster-like topological defects are connected by a wavy pattern, due to the chiral nature of the nematic field. (b)As activity is increased, the energy injected is able to relocate the two defects in a new configuration. The competing mechanism between active injection and elastic relaxation leads the droplet to rotate with the two defects preceding around each other in a fan-like fashion. Astonishingly, this leads the droplet to propel straight.

## 4. SPONTANEOUS FLOW AND DROPLET MOBILITY

In this Section we will present some results of our study concerning the properties of active fluids, obtained by integrating the nemato-hydrodynamic equations introduced in Section 2.
It is common experience that a fluid not subject to any force relaxes into its equilibrium configuration characterized by the absence of flows. Nevertheless, active fluids are characterized by continuous injection of energy due to the action of the swimmers on the surrounding fluids. In accordance of the intensity of the forcing, active fluids may settle in three different states. When the rate at which activity injects energy in the system is low compared with viscous effects, energy is locally dissipated and the fluid relaxes into a *quiescent* state, where the LC is globally ordered and flows are almost suppressed, with no significant difference from the passive counterpart.
If the intensity of active forcing is raised, the liquid crystal which models the active components, becomes unstable and settle in a wavy pattern (panel a of Fig. 4). Bending deformations of the nematic pattern act as a source of momentum (since the active stress tensor is proportional to the gradients of the nematic tensor $Q_{\alpha\beta}$), thus generating a self-sustained and stationary flow (see the superimposed velocity field in Fig. 4). This non-equilibrium steady state, developing when energy injection rate is comparable with dissipation rate, is commonly addressed as *spontaneous flow* [1]. By further increasing activity, the liquid crystal deformations strengthen, generating more and more intense flows that drive the system into a chaotic state (panel b of Fig. 4). The typical flow pattern resembles those observed in turbulent low viscosity media. This is somehow surprising, since most active fluids are considerably viscous and the development of a turbulent state is highly unexpected. Moreover, this state is characterized by the formation of topological defects. These are regions where the LC settle in vortical patterns so that the alignment direction cannot be properly defined.
The richness of behaviors which characterizes active fluids and the possibility to tune the response of the system by varying a single control parameter had brought the scientific community to think that such fluids may be exploited to design new smart materials and Lab-on-chips. To this aim, active fluids must be opportunely confined and dispersed in a fluidic (passive) environment.

For this reason, much effort has been spent in the last decade to understand and control the behavior of droplets of active material suspended in an isotropic fluid [19]. Analogously to what happens for a fully nematic suspensions, even in this case the behavior of an active droplet is characterized by a quiescent state, spontaneous flow and an active turbulent regime which can be selected by controlling the intensity of active forcing. Importantly, the spontaneous flow state may result either in the rotational motion or in the self-propulsion of the droplet itself. Recently we have analyzed [15], by means of lattice Boltzmann simulations, a system consisting of a droplet of extensile cholesteric liquid crystal to exploit the important relation between activity and chirality –a feature common to many biological filaments, such as microtubules or acto-myosin, that are implemented in actual experiments on active matter.

By fixing the pitch of the helix of the liquid crystal as half of the radius of the droplet, we systematically varied the intensity of the activity parameter. For small active forcing the droplet settles in the quiescent state, characterized by the absence of flows. The LC, tangentially anchored to the interface of the droplet, forms two antipodal aster-like topological defects, connected by a wavy pattern, due to the chiral nature of the nematic field (panel a of Fig. 5).

As activity is increased, the energy injected is able to relocate the two defects in a new configuration, as shown in panel b of Fig. 5. Such state is commonly known as *Frank-Pryce* structure and is typically observed for higher cholesteric power. Here, the energy injection favors the deformation of the LC pattern inside the droplet, strengthening the frustration. Moreover, the competing mechanism between active injection and elastic relaxation leads the droplet to rotate. The two defects start preceding around each other in a fan-like fashion, as indicated by the arrow in Fig. 5b. Astonishingly, the rotational motion is also accompanied by the drift of the droplet itself. This propulsive mechanism is peculiar of cholesteric droplets and cannot be observed in nematic LC (that do not develop helicoidal structures). Indeed, the asymmetric structure of defects on the droplet surface is a necessary condition for the droplet to move straight, with a drift velocity proportional to the intensity of the active forcing.

Finally, when activity is increased over a certain threshold, the droplet sets into a chaotic state characterized by the proliferation of defects both on the surface of the droplet and in its interior (not shown here).

## 5. CONCLUSIONS

In this article we presented a brief overview of the active gel theory that has been widely used to study fluidic active matter. After presenting the dynamical equations, we presented the lattice Boltzmann algorithm that we made use of to integrate the dynamics of the system. Moreover, to explore three-dimensional systems a parallel approach is compulsory to overcome issues due to long simulation times and high memory requirements. We presented the idea beyond the parallelization scheme, that we carried out by means of MPI, and a scaling test performed on three different computing infrastructures. In the second part of the article we presented some fundamental properties of active fluids. We identified three different dynamical states which can be selected by operating on a single control parameter, namely the intensity of active forces. For low energy injection a quiescent flowless state sets up. As activity is increased, the system enters a state addressed as *spontaneous flow* in which the liquid crystal develops stationary bending instabilities that act as a source of momentum thus generating stationary self-sustained flows.

By further increasing activity, the active fluid develops chaotic structures that span the system. The local flow generated by the topological defects plays an important role in confined systems. In this case, the dynamics of the defects can be controlled, so to generate macroscopic coherent flows which may lead to the development of motility properties to be exploited in the design of novel devices. In our numerical study we considered the case of a droplet of active cholesteric LC suspended in an isotropic fluid. We showed that the feedback interaction between active injection and elasticity effects due to chirality leads to the linear propulsion of the droplet thanks to the rearrangement of topological defects.